%% file: version2hr.tex
\documentclass[prb,aps,epsfig,twocolumn,showpacs]{revtex4}

\usepackage{epsfig,amssymb,amsmath,latexsym}

\input superqpumpmacro

\begin{document}

\textheight=23.8cm

\title{Quantized charge pumping in superconducting double barrier
structure : Non-trivial correlations due to proximity effect}

\author{Arijit \surname{Saha}}
\email[Email: ]{arijit@hri.res.in}
\affiliation{Harish-Chandra
Research Institute, \\ Chhatnag Road, Jhusi, Allahabad 211019,
India}
\author{Sourin \surname{Das}}
\email[Email: ]{sourin.das@weizmann.ac.il}
\affiliation{
Department of Condensed Matter Physics, \\
Weizmann Institute of Science, Rehovot 76100, Israel }

\date{\today}

\begin{abstract}
We consider quantum charge pumping of electrons across a
superconducting double barrier structure in the adiabatic limit. The
superconducting barriers are assumed to be reflection-less so that
an incident electron on the barrier can either tunnel through it or
Andreev reflect from it. In this structure, quantum charge pumping
can be achieved {\textsl{(a)}} by modulating the amplitudes,
$\Delta_1$ and $\Delta_2$, of the gaps associated with the two
superconductors or alternatively, {\textsl{(b)}} by a periodic
modulation of the order parameter phases, $\phi_1$ and $\phi_2$ of
the superconducting barriers. In the former case, we show that the
superconducting gap gives rise to a very sharp resonance in the
transmission resulting in quantization of pumped charge, when the
pumping contour encloses the resonance. On the other hand, we find
that quantization is hard to achieve in the latter case. We show
that inclusion of weak electron-electron interaction in the quantum
wire leads to renormalisation group evolution of the transmission
amplitude towards the perfectly transmitting limit due to interplay
of electron-electron interaction and proximity effects in the wire.
Hence as we approach the zero temperature limit, due to
renormalisation group flow of transmission amplitude we get
destruction of quantized pumped charge. This is in sharp contrast to
the case of charge pumping in a double barrier through a Luttinger
liquid where quantized charge pumping is actually achieved in the
zero temperature limit.
\end{abstract}

\maketitle

\section{\label{sec:one} Introduction}

The phenomena of quantum charge pumping corresponds to a net flow of
\dc current between different electron reservoirs (at zero bias)
connected via a quantum system whose system parameters are
periodically modulated in time~\cite{brouwer,bpt}. The zero bias
current is obtained in response to the time variation of the
parameters of the quantum system which explicitly break the time
reversal symmetry. It is worth mentioning that breaking of the
time-reversal symmetry is necessary but not a sufficient condition
in order to get net pumped charge in unit cycle. For obtaining a net
pumped charge, parity or spatial symmetry must also be broken.
Within a scattering approach, if the time period of modulation of
the scattering system parameters is much larger than the time the
particle spends inside the scattering region, adiabatic limit is
reached. In this limit, the pumped charge in a unit cycle becomes
independent of the pumping frequency. This is referred to as
``adiabatic charge pumping''. Experimentally charge pumping has been
observed in mesoscopic systems involving quantum dots and carbon
nanotubes~\cite{marcus,leek,buitelaar}. Ofcourse one has to be very
careful in interpreting the experimentally observed pumped charge as
it can be faked by rectification effects as was pointed out by
Brouwer~\cite{brouwer1}.

\begin{figure}[htb]
\begin{center}
\includegraphics[width=8cm,height=5cm]{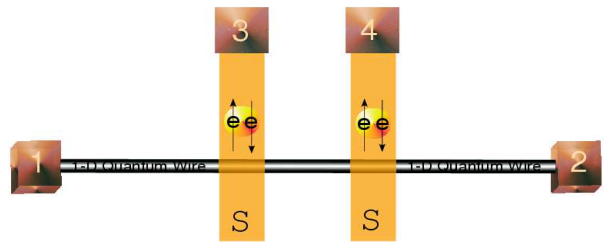}
\vskip -2.0cm
\caption{A one-dimensional quantum wire ({\sl{e.g.}} carbon
nanotube) connected to two reservoirs, labelled by 1 and 2. The
two thin strips on the wire depict layers of superconducting
material deposited on top of the wire. The superconducting strips
are connected to contacts labelled 3 and 4.}
\label{geometry}
\end{center}
\end{figure}

In the recent years, there has been an upsurge of research interest
in exploring the effects due to inclusion of electron-electron
($\ed-\ed$) interaction on the pumped charge~\cite{sharma,
brouwer1,governale,sela1,silva,andrei,amit}. In this article, we
explore the effect of inter-electron interaction on the charge
pumped across a superconducting double barrier (\sdbd)
system~\cite{morpurgo} in the context of one-dimensional (\odd)
quantum wire (\qwd). Pumping of free electrons across \od quantum
well was studied earlier in
Refs.~\onlinecite{das2003bdr,wohlman,saha}, where using Brouwer's
formula~\cite{brouwer}, it was shown that the pumped charge can be
expressed as a sum of two contributions, viz., a dissipative part
and a quantized topological part, the latter being independent of
the details of the pumping contour~\cite{aleiner,levinson}. The
dissipative part was found to be proportional to the conductance
through the system on the pumping contour in the parameter space
while the topological part was non-zero only if the pumping contour
enclosed a resonance. Hence in order to obtain quantized pumped
charge, one needs to reduce the dissipative part as much as
possible. This is very naturally achieved if one considers pumping
through a quantum well in a \od interacting electron
gas~\cite{sharma,das2004dr} (Luttinger liquid) as in this case
interaction correlations make the resonance very sharp thereby
reducing the conductance on the contour enclosing the resonance to
vanishingly small values in the zero temperature limit. This leaves
behind a quantized topological part. The pumped charge was shown to
converge to a quantized value asymptotically. This was obtained
using a perturbative approach for the case of a weakly interacting
electron gas followed by ``Poor-man's scaling approach"~\cite{yue}.
In this article, we show that the presence of inter-electron
interaction in the wire leads to nontrivial scattering processes due
to proximity effects which leads to a power law reduction of the
pumped charge from the quantized value (as opposed to enhancement)
in the adiabatic limit as one lowers the temperature. Quantum charge
pumping using various setups involving superconductor has been a
topic of major interest in recent
 past~\cite{wangsup,fzhou,blaauboer,governale1,taddei1,taddei2,kopvin,wangwang,
wangwang1,morpurgoPRL}. Specifically we consider pumping of
electrons (in the adiabatic limit) across a \sdb structure, as
depicted in Fig.~\ref{geometry}. Till date no experiment has been
carried out in the context of charge pumping for the case of
superconducting barrier. Experimentally it might be possible to
design a \sdb structure by depositing thin strips of superconducting
material on top of a single ballistic \qw (like carbon nanotubes) at
two places, which can induce a finite superconducting gap in the
barrier regions of the \qw as a result of proximity of the
superconducting strips. In our simple-minded theoretical modelling
of the system we assume that the superconducting barrier (\scbd) to
be reflection-less so that an incident electron on the barrier can
either tunnel through it or Andreev reflect from it. Within the
simplified theoretical model, we explore two scenarios to achieve
quantization of pumped charge $-$ {\textsl{(a)}} by periodic
modulation of amplitudes $\Delta_1$ and $\Delta_2$ of the gap at the
two \scb or alternatively, {\textsl{(b)}} by periodic modulation of
the order parameter phases $\phi_1$ and $\phi_2$ associated with the
two \scbd. For free electrons in the \qw we show that in the
$\Delta_1-\Delta_2$ plane, there is an isolated sharp resonance
point in transmission probability across the \sdb structure. On the
other hand transmission probability across the double barrier has a
line of sharp resonances in the $\phi_1-\phi_2$ plane. As mentioned
earlier, in order to obtain quantized pumped charge, the
transmission on the pumping contour should be as small as possible.
When we consider $\Delta_1$ and $\Delta_2$ as the pumping
parameters, we can always choose a pumping contour which completely
encloses the isolated resonance and hence it is possible to achieve
quantization of charge if the resonance is sharp enough. However, in
the $\phi_1-\phi_2$ plane, we have a line of resonances. Any closed
contour enclosing the resonances will surely cross the resonance
line at least twice thereby increasing the dissipative part and
consequently resulting in destruction of quantization of pumped
charge. Interestingly enough, inclusion of weak $\ed-\ed$
interaction in the wire results in a \rg flow of the transmission
amplitude towards perfectly transmitting limit due to proximity
induced effect on the interacting electrons in the \qw as
%
%
we lower the energy scale such as temperature. Hence the sharpness of
resonance is lost due to \rg enhancement of transmission through individual
\scb resulting in complete destruction of quantized charge pumping
as we go down in temperature. It is worth noticing that the
consequence of inclusion of correlations due to $\ed-\ed$
interaction is just opposite here with respect to the case
of double barrier in a Luttinger liquid~\cite{das2004dr}.\\
This article is organised as follows. In Sec.~II, we discuss the
modelling of \sdb in a \od \qw and calculate the transmission and
Andreev reflection (\ard) amplitudes of the system. In Sec.~III, we
discuss the renormalisation group (\rgd) flow for transmisson and
\ar for the \scbd. In Sec.~IV, we discuss our \rg scheme for the
\sdb and calculate the pumped charge. In the end, we discuss our
results and give conclusions in Sec.~V.

\section{\label{sec:two}Superconducting barrier}
Quantum transport in \scb structure was considered in past in
Ref.~\onlinecite{morpurgo}. Here we consider a very similar set-up
comprising of a ballistic \od \qw with two short, but finite
superconducting patches as shown in Fig.~\ref{geometry}. Here,
$\Delta^{(i)}$ is the pair potential on the two patches ($i$ refers
to the index of the strip).
Following Ref.~\onlinecite{morpurgo}, we use the Bogolubov$-$de
Gennes (\bdgd) equation~\cite{degennes,blonder} to calculate the
transmission amplitude $t_{ee}^{(i)}$ and the \ar amplitude,
$r_{eh}^{(i)}$, where $i$ is the barrier index. The space dependance
of the order parameter (which also acts as the scattering potential)
for the incident electron can be expressed as
\begin{eqnarray} V(x) &=& \Delta^{(i)} e^{i \phi_1} \Theta(x)
\Theta(-x+a) + \Delta^{(i)}
e^{i\phi_2} \non\\
&&  \Theta[x-(a+L)] \Theta[-x+(2a+L)]
\end{eqnarray} where, $a$ is the width of the \scb and $L$ is the
distance between the two barriers. \\
%
%
Hence the \bdg equations can be written as, \bea Eu_{+} &=&
\left[\dfrac {-\hbar^{2} \nabla^{2}} {2m} + V(x) - \mu \right]u_{+}
+ \Delta u_{-} \eea \bea Eu_{-} &=& \left[\dfrac {\hbar^{2}
\nabla^{2}} {2m} - V(x) + \mu \right]u_{-} + \Delta^\star u_{+} \eea
%
Solving the \bdg equation in the normal and superconducting regions and
matching the solution at $x=0$ and $x=a$, we get
\bea t_{ee}^{(i)}
&=& \dfrac{e^{ik^+a}(u_+^2-u_-^2)}{u_+^2 - u_-^2
e^{i(k^+-k^-)a}}\,;
 \quad t_{hh}^{(i)} = \dfrac{e^{-ik^-a}(u_+^2-u_-^2)}
{u_+^2 - u_-^2 e^{i(k^+-k^-)a}}
\non
\eea \bea
 r_{eh}^{(i)} &=&
\dfrac{u_+ u_- e^{-i\phi_{i}}(1-e^{i(k^+-k^-)a})}{u_+^2 - u_-^2
e^{i(k^+-k^-)a}}
\non\\
r_{he}^{(i)} &=& \dfrac{u_+ u_-
e^{i\phi_{i}}(1-e^{i(k^+-k^-)a})}{u_+^2 - u_-^2 e^{i(k^+-k^-)a}} \eea
\noindent where
$\hbar k^\pm = \sqrt{2m(E_F \pm (E^2-\Delta^{2(i)})^{1/2})}$,
$u_{\pm}=\frac {1}{\sqrt{2}}
[(1\pm(1-(\Delta^{(i)}/E)^{2})^{1/2})]^{1/2}$
Here, $t_{ee}^{(i)}$,  $t_{hh}^{(i)}$, $r_{eh}^{(i)}$,
$r_{he}^{(i)}$ are the transmission and \ar amplitudes. $m$ is the
effective mass of the electron in the wire, $E_F$ is the Fermi
energy for the electrons in the superconducting region, and $E$ is
the Fermi energy of electrons in the normal region of the \qwd,
measured with respect to $E_F$. Hence the scattering matrix for
the single \scb problem for an incident electron or hole is given
by
\bea S_{e} =
\begin{vmatrix} ~r_{eh}^{(i)} & t_{ee}^{(i)}~\\
~t_{ee}^{(i)}& r_{eh}^{(i)}~\\
\end{vmatrix} \quad \And \quad  S_{h} =
\begin{vmatrix} ~r_{he}^{(i)}& t_{hh}^{(i)}~\\
~t_{hh}^{(i)}& r_{he}^{(i)}~\\
\end{vmatrix}
\label{smatsingle} \eea
Using the $S$-matrix given by Eq.~\ref{smatsingle}, we obtain the
effective $S$-matrix for the double barrier system~\cite{sdatta}. We
assume particle-hole symmetry, hence $t_{ee}=t_{hh}$ and $r_{eh}=r_{he}$.
The net transmission and net \ar amplitude through the double barrier are
\bea T_{ee} = \dfrac{t_{ee}^{(1)} t_{ee}^{(2)}
e^{i q^+ L}}{1-r_{eh}^{(2)}
r_{he}^{(1)} e^{i (q^+-q^-) L}} \non\\
R_{eh} = r_{eh}^{(1)} + \dfrac{t_{ee}^{(1)} r_{eh}^{(2)}
t_{hh}^{(1)} e^{i (q^+-q^-)L}}{1-r_{eh}^{(2)} r_{he}^{(1)} e^{i
(q^+-q^-) L}} \eea \noindent where $\hbar q^{\pm}=\sqrt{2m(E_{F} \pm
E)}$.
In order to obtain quantization of pumped charge, we choose to operate in
the sub-gap regime \ie, $E < \Delta$. 
In this regime, $|T_{ee}|^2$ has sharp resonances at discrete values of
$E/\Delta$ for a given value of $\phi_1-\phi_2$~\cite{morpurgo}.
These resonances result from multiple \ar of electron to hole and
vice-versa inside the double barrier.
%

\section{\label{sec:three} \wirg study of junctions}
We study the effects of inter-electron interactions in the wire on
the $S$-matrix characterizing the superconducting barrier using the
\rg method introduced in Ref.~\onlinecite{yue}, and the
generalizations to multiple wires in
Refs.~\onlinecite{lal,das2004drs}. The basic idea of the method is
as follows. The presence of back-scattering (reflection) induces
Friedel oscillations in the density of non-interacting electrons.
Within a mean field picture for weakly interacting electron gas, the
electron not only scatters off the potential barrier but also
scatters off these density oscillations with an amplitude
proportional to the interaction strength. Hence by calculating the
total reflection amplitude due to scattering from the scalar
scatterer and from the Friedel oscillations created by the
scatterer, we can include the effect of $\ed-\ed$ interaction in
calculating transport. This can now be generalized in a similar
spirit to the case where there is, besides non-zero reflection also
non-zero \ar which turns an incoming electron into an outgoing hole
due to proximity effects as done in Ref.~\onlinecite{jap3} and then
generalized to multiple wire superconducting junction in
Ref.~\onlinecite{das2007drsahaprb}.

The fermion field on each wire can be written as,
\beq \psi_{is} (x) = \Psi_{I\,is}(x)\, e^{i\,k_F\,x} \,+\,
\Psi_{O\,is}(x)\, e^{-i\,k_F\,x} \eeq
where $i$ is the wire index, $s$ is the spin index which can be
$\up,\dn$ and $I,O$ stands for outgoing or incoming fields. Note
that $\Psi_{I}(x) (\Psi_O (x))$ are slowly varying fields on the
scale of $k_F^{-1}$ and contain the annihilation operators as well
as the slowly varying wave-functions.
%
%
The expectation values for the density
$\scxone{\Psi_{is}^\dagger\Psi_{is}}$ gives (dropping the wire index),

\bea \scxone{\psi_{O\,\up}^\dagger \psi_{I\,\up}} ~=~
\scxone{\psi_{O \,\dn}^\dagger \psi_{I\,\dn}} ~=~ \frac
{i\, r^\star}{4 \pi x}\\
\non \\\And \quad \scxone{\psi_{I\,\up}^\dagger \psi_{O \,\up}} ~=~
\scxone{\psi_{I\,\dn}^\dagger \psi_{O\,\dn}} ~=~ \frac {-i \,r} {4
\pi x}. \label{boguliobov} \eea Hence, besides the density, the
expectation values for the pair amplitude
$\scxone{\Psi_{is}^\dagger\Psi_{is}^\dagger}$ and its complex
conjugate $\scxone{\Psi_{is}\Psi_{is}}$ are also non-zero and are
given by (dropping the wire index)
\bea \scxone{\psi_{O\,\up}^\dagger \psi_{I\,\dn}^\dagger} ~=~ -
\scxone{\psi_{O \,\dn}^\dagger \psi_{I\,\up}^\dagger} ~=~ \frac
{-i\, r_{A}}{4 \pi x}\\
\non \\\And \quad
\scxone{\psi_{O\,\up} \psi_{I \,\dn}} ~=~ - \scxone{\psi_{O\,\dn}
\psi_{I\,\up}} ~=~ \frac {-i \,r_{A}^\star} {4 \pi x}. \label{boguliobov}
\eea
So, we see that the Boguliobov amplitudes fall off as $1/x$
just like the normal density amplitudes.\\
%
We now allow for short-range density-density interactions between
the fermions
\beq
\hmi = \frac{1}{2} \, \int dx \,dy \, \left(\sum_{s\,=\,\up,\dn}
\rho_{s}\right) \,V(x-y)\,\left(\sum_{s\,=\,\up,\dn}
\rho_{s}\right)
\eeq
to obtain the standard four-fermion interaction Hamiltonian for
spin-full fermions as
\bea
\hmi & = & \int dx \Big[ g_1 \Big(\psiiudg  \psioudg \psiiu \psiou
\,+\, \psiiddg \psioddg \psiid \psiod
\nonumber\\
&+&   \psiiudg \psioddg \psiid \psiou \,+\, \psiiddg \psioudg \psiiu
\psiod\Big)
\nonumber \\
&+&  g_2  \Big(\psiiudg \psioudg \psiou \psiiu
+ \psiiddg \psioddg  \psiod \psiid
\nonumber \\
&+&  \psiiudg \psioddg \psiod \psiiu  \,+\,
\psiiddg \psioudg \psiou  \psiid\Big)\Big]
\nonumber \\
&& \eea where $g_1$ and $g_2$ are the interaction parameters \cite{solyom}.

The effective Hamiltonian can be derived using a Hartree$-$Fock (\hfd)
decomposition of the interaction Hamiltonian. The charge conserving \hf
decomposition leads to the interaction Hamiltonian
(normal) of the following form on each half wire,
\bea
\hmi^N &=& \dfrac{-i(g_2-2g_1)}{4\pi}  \int_0^\infty \dfrac{dx}{x}
\Big[r^\star \left(\psiiudg \psiou + \psiiddg \psiod \right)
\nonumber \\
&-& r \left(\psioudg \psiiu + \psioddg \psiid \right)\Big] \eea
(We have assumed spin-symmetry \ie~ $r_{\up} = r_{\dn} = r$.) This
has been derived earlier~\cite{lal}. Using the same method, but
now also allowing for a charge non-conserving \hf decomposition we get
the (Andreev) Hamiltonian
\bea
\hmi^A &=& \dfrac{-i(g_1+g_2)}{4\pi}  \int_0^\infty  \dfrac
{dx}{x} \Big[-r_{A}^{\star}\big(\psiiudg \psioddg +
 \nonumber \\
&& \psioudg \psiiddg \big)
 + r_{A}\left( \psiod\psiiu + \psiid\psiou \right) \Big]\eea

The $\ed-\ed$ interaction induced amplitude to go from an incoming
electron wave to an outgoing electron wave under $e^{-i{\hmi^N}t}$
(for electrons with spin) is given by \cite{lal}
\beq {-\alpha \, r_{s} \over 2} \, \ln (kd) \label{Friedeln} \eeq
where $\alpha = (g_2-2g_1) / 2\pi\hbar v_F $ and $d$ is the short
distance cut-off for the \rg flow. Analogously, the amplitude to go
from an incoming electron wave to an outgoing hole wave under
$e^{-i{\hmi^A} t}$ is given by~\cite{jap3}

\beq {\alpha^\prime \,r_{A} \over 2} \, {\ln (kd)} \label{FriedelA} \eeq
where $\alpha^\prime = (g_1+g_2)/ 2 \pi  \hbar  v_F $.\\
These logarithmic corrections to the bare reflection amplitude and
the \ar amplitude can be summed up using a ``Poor-man's scaling
approach''~\cite{anderson} which finally leads a \rg equation for
$r$ and $r_{A}$.

\section{\label{sec:four}Renormalisation group scheme and the pumping formula}
We include the effects due to proximity of superconductor and
$\ed-\ed$ interaction in the wire via a \rg approach developed very
recently~\cite{das2007drsahaprb} for the case of \od normal
metal$-$superconductor$-$normal metal (\nsnd) junction. As we are
only interested in the coherent regime ($L_T
>> L$, $L_T$ is the thermal length), we can effectively treat the \sdb
system (\nsnsn junction) as a single barrier (\nsn junction) as far
as \rg is concerned.

\begin{figure*}[htb]
\begin{center}
\includegraphics[width=7cm,height=7cm]{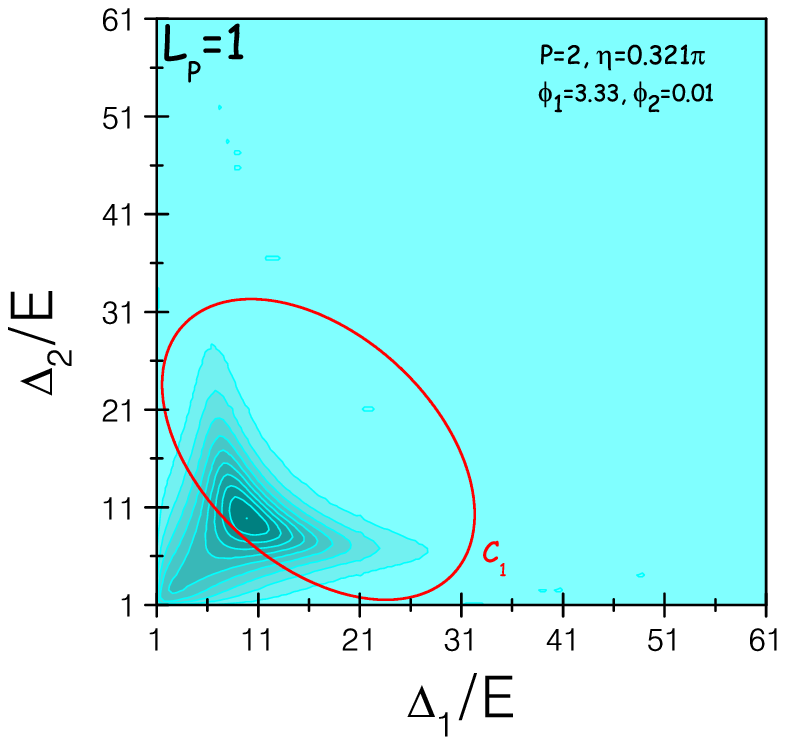}
\hskip 0cm
\includegraphics[width=9cm,height=7cm]{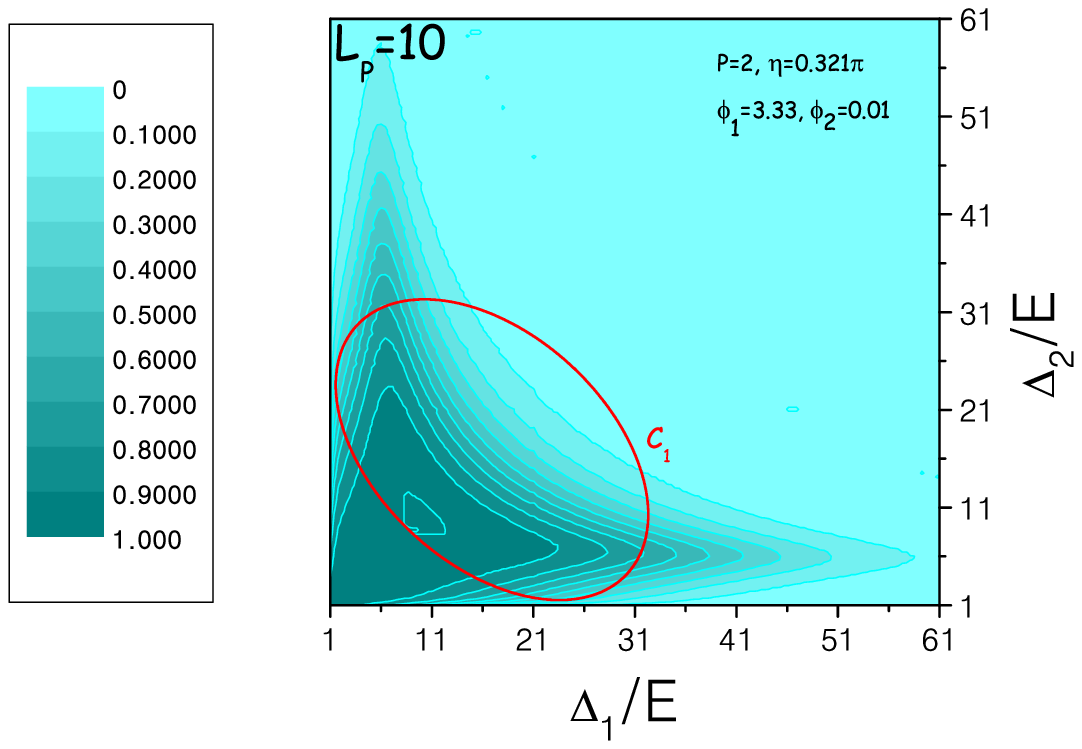}
\hskip 0cm \caption{Contours of transmission probability, $|T_{ee}|^{2}$
in the $\Delta_1-\Delta_2$ plane at two different values of length
scale, $L_P=1$ and $L_P=10$, at which the \rg flow is cut-off for
values of $V(0)=0.1$ and $V(2k_{F})=0.1$. The red ellipse $C_1$
represents the pumping contour.} \label{contamp}
\end{center}
\end{figure*}
\begin{figure*}[htb]
\begin{center}
\includegraphics[width=7cm,height=7cm]{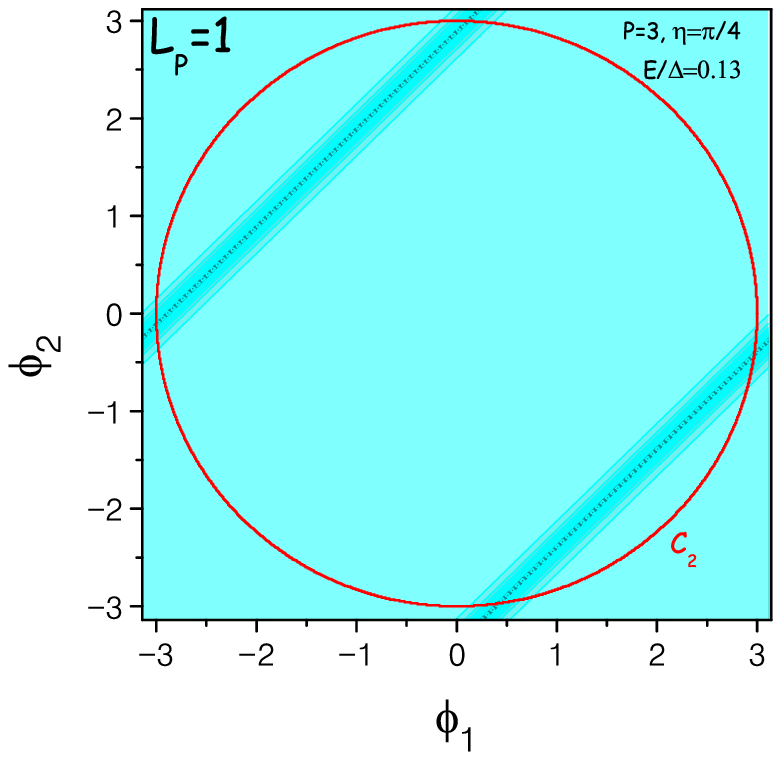}
\hskip 0cm
\includegraphics[width=9cm,height=7cm]{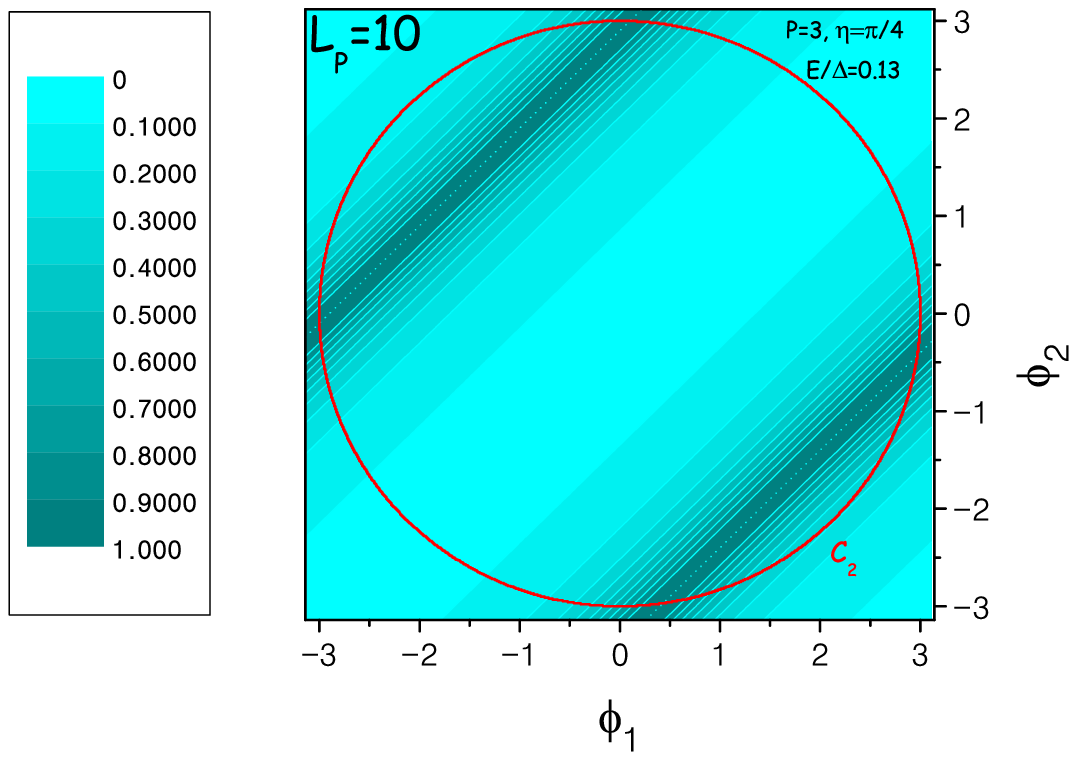}
\caption{Contours of transmission probability, $|T_{ee}|^{2}$ in the
$\phi_1-\phi_2$ plane at two different values of length scale,
$L_P=1$ and $L_P=10$ at which the \rg flow is cut-off for values of
$V(0)=0.1$ and $V(2k_{F})=0.1$. The red circle $C_2$ represents
the pumping contour.}
 \label{contph}
\end{center}
\end{figure*}
Hence the effective two-channel $S$-matrix  for this double barrier
system can be written as

\beq S =
\begin{vmatrix}
~|R_{eh}|e^{i\theta} & |T_{ee}| e^{i\phi} ~\\
~|T_{ee}| e^{i\phi} & |R_{eh}| e^{i\theta'}~
\end{vmatrix}
\label{two} \eeq
where all the amplitudes and phases associated with the matrix
elements are functions of the time-varying parameters, $V_i (t) =
V_0 + P \cos(\omega t + (-1)^{i-1} \eta)$ where $i=1,2$ stands for
the  barrier index. $V_i = \Delta_i$ and $V_i = \phi_i$ are the two
possible pumping parameters. The reflection coefficients are not the
same (phases can differ) because the time-varying potentials
explicitly violate parity. In principle the $S$-matrix also violates
time-reversal invariance. But since in the adiabatic approximation,
we are only interested in instantaneous hamiltonian. Note that the
instantaneous $S$-matrix can mimic a time-reversal symmetric
$S$-matrix.

Using the modified Brouwer's formula~\cite{morpurgoPRL}, the pumped
charge can directly be obtained from the parametric derivatives of
the $S$-matrix elements. It is worth mentioning that even though
Brouwer's formula is valid for non-interacting electron system, we
are able to use it here because effects due to interactions in the
wires can be taken care of by the renormalization of the bare
$S$-matrix obtained for the free-electron case.

For single channel $S$-matrix, we
have \bea {\cal Q} &=& {e\over 2 \pi} \int_0^\tau dt~
{\mathrm{Im}}~ \Bigg[ -{\partial S_{11}\over
\partial V_1}S_{11}^\star {\dot V_1} + {\partial S_{12}\over \partial
V_1}S_{12}^\star {\dot V_1}
\nonumber \\
&&~~~~~~~~~~\quad-{\partial S_{11}\over \partial V_2}
S_{11}^\star{\dot V_2} + {\partial S_{12}\over \partial V_2}
S_{12}^\star{\dot V_2}\Bigg]\label{pc} \eea where $S_{ij}$ denote
the elements of the $S$-matrix. Note the negative sign in the
above expression, which results from the fact that $S_{11}$
corresponds to conversion of an electron into a hole. Thus, the
pumped charge is directly related to the amplitudes and phases
that appear in the $S$-matrix. Inserting Eq.~\ref{two} in
Eq.~\ref{pc},
\beq {\cal Q} = {e\over 2\pi} \int_0^\tau \left[~
{\dot \theta} -
G(t)({\dot \theta} +{\dot \phi})~\right] dt
\label{pc2} \eeq
Here $G(t)=|T_{ee}(t)|^{2}$ is the instantaneous two terminal linear
conductance (labelled by $1,2$ in Fig.~\ref{geometry}), in units of
$2e^2/h$. The first term on the RHS in Eq.~\ref{pc2} is clearly
quantized since $e^{i\theta}$  returns to itself at the end of one
cycle. So the only possible change in $\theta$ in a period can be in
integral multiples of $2 \pi$ \ie, $\theta(\tau) \rightarrow
\theta(0) + 2\pi n$ ($n \to$ integer). The second term is the
`dissipative' term which prevents the perfect quantization. The
second term is directly proportional to the two terminal
Landauer$-$Buttiker conductance for the system on the pumping
contour. The relative sign between $\dot \theta$ and $\dot \phi$ in
the expression for pumped charge in Eq.~\ref{pc2} originates from
the \ar process, which converts an electron to a hole. This is in
contrast to what has been found for the normal double barrier
problem~\cite{das2004dr}. For a reflection-less junction, the basic
idea of the \rg method is as follows. The presence of a
superconductor induces a finite yet weak pair potential in the \qw
resulting in scattering of incoming electrons to outgoing holes
(Andreev processes) in the wire, away from the junction. Hence by
calculating the total \ar amplitude, due to scattering from the \nsn
junction and the (weak) pair potential in the wire perturbatively in
interaction strength and followed by ``Poor-man's scaling" approach,
we obtain the \rg equation for the elements of the effective
$S$-matrix of the \sdb structure in the coherent regime ($L_T > L$).
\begin{figure}[htb]
\begin{center}
\includegraphics[width=7cm,height=5.5cm]{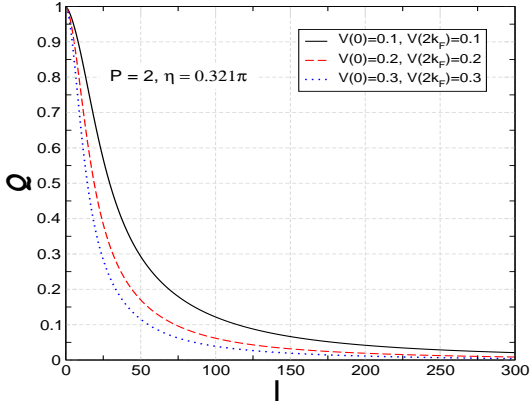}
\caption{Pumped charge $\cal Q$, for pumping in $\Delta_1-\Delta_2$ plane,
is shown as a function of the dimensionless parameter $l$ where
$l=ln(L_{P}/d)$ and $L_{P}$ is either $L_{T}=\hbar v_{F}/k_{B} T$ at zero
bias or $L_{V}=\hbar v_{F}/eV$ at zero temperature and $d$ is the short
distance cut-off for the \rg flow. The three curves correspond to three
different values of $V(0)$ and $V(2k_{F})$.}
 \label{figpumpchargeamp}
\end{center}
\end{figure}
So, the entries of $S$-matrix therefore become functions of the length
scale $L_{P}$ due to the \rg flow.
The \rg flow can also be considered to be a flow in the
temperature since the length scale $L_{P}$ can be converted to a
temperature scale using the thermal length $L_T = {\hbar v_F}/{(k_B
T)}$. Hence, the \rg flow has to be cut-off by either $L_T$, or
the system size $L_S$, whichever is smaller~\cite{das2004drs}.

Without loss of generality, we can calculate the renormalized
$S$-matrix at different length scales or equivalently at different
temperatures at any point on the pumping contour.
Hence, to avoid unnecessary complications arising due to the \rg flow of
phases associated with $S$-matrix elements
($\theta,\theta',\phi$), we choose to calculate the \rg flow of the
$S$-matrix when the barriers are symmetric. This symmetry leads to
vanishing of the \rg flow of the phases hence making the calculation
algebrically simple.

The \rg flow of the normal transmission (and \ard) amplitudes and
phases are~\cite{das2007drsahaprb}
\bea {d|T_{ee}|\over dl} &=& \alpha' |T_{ee}| (1-|T_{ee}|^2) \quad
\And \quad \label{rgtone}
 {d\phi \over dl} = 0
\label{rgtrans}
\non
\eea
\bea {d|R_{eh}|\over dl} &=& -{\alpha'\over 2} |R_{eh}|
[1-|R_{eh}|^2
-|T_{ee}|^2\cos 2(\phi-\theta)] \label{rgone} \non\\
{d\theta \over dl} &=& {\alpha'\over 2}|T_{ee}|^2 \sin
2(\phi-\theta) \label{rg} \eea
Here $l=ln(L_{P}/d)$ where $d$ is the short distance cut-off for the
\rg flow and  we have considered the fully symmetric case,
\ie~$\theta = \theta'$. Unitarity of the $S$-matrix in Eq.~\ref{two}
implies that $\phi - \theta = \pi/2 + 2n\pi$ ($n \to$ integer). This
simplifies the equations for \rg flow for the \ar amplitude and
phase,
\bea {d|R_{eh}|\over dl} = - \alpha' |R_{eh}|
\left(1-|R_{eh}|^2\right) \quad \And \quad \label{rgtone2}
 {d\theta \over dl} = 0 \label{rgtrans2} \eea
\noindent  Here, $\alpha^\prime = (g_2 + g_1)/2\pi \hbar v_F$
where $g_1,g_2$ are the running coupling constants whose bare
values are set by $g_1(L_{P}=d) = V(2k_F)$ and $g_2(L_{P}=d) =
V(0)$; $V(x)$ being the inter-electron interaction potential.
We now integrate the \rg equation for $T_{ee}$ complimented by the
\rg flow of $g_1$ and $g_2$~\cite{das2007drsahaprb} to obtain the
$L_{P}$ dependence of $T_{ee}$ as
\bea T_{ee}(L_{P}) &=& \non\\ &&
\!\!\!\!\!\!\!\!\!\!\!\!\!\!\!\!\!\!\!\!\!\!\!\!\!\!\!\!\!\!\!\!\!\!\!
\!\!\!\!\!\!\!\!\!\!\!\!\!\!\!
 \frac{T_{ee}^{0} \left[\left(1 + 2\alpha_{1} \ln
\frac{L_{P}}{d}
\right)^\frac{3}{2}\left(\frac{d}{L_{P}}\right)^{-(2\alpha_{2}
-\alpha_{1})}\right]}{R_{eh}^{0}+T_{ee}^{0}\left[\left(1+2\alpha_{1}
\ln
\frac{L_{P}}{d}\right)^\frac{3}{2}\left(\frac{d}{L_{P}}\right)
^{-(2\alpha_{2}-\alpha_{1})}\right]} \label{ten}  \eea
Here $T_{ee}^0$ and $R_{eh}^0$ are the values of $T_{ee}$ and
$R_{eh}$ at length-scale $L$ and $\alpha_1 = V(0)/2\pi\hbar v_F$ and
$\alpha_2 = V(2k_F)/2\pi\hbar v_F$. There are two points worth
mentioning here :
{\textsl{(a)}} the transmission increases with increasing $L_{P}$
which is a consequence of the fact that the proximity effect due
to superconductor induces an effective attractive interaction
between the electrons, hence rendering the (Andreev)
back-scattering an irrelevant operator, and {\textsl{(b)}} the
expression for $T_{ee}(L_{P})$ is not in the form of a pure power
law even at $T_{ee}^0$ $\rightarrow 0$ limit, as is expected from
Luttinger Liquid physics because of the \rg flow of the $g_1,g_2$
parameters. Also, it is important to note that we take the
short-distance cut-off $d$ to be the distance between the two
barriers ($L$) since this is the length scale at which we glued
the two barriers to a single barrier as far as \rg is concerned.
Using this, we can obtain the scaling behavior of the pumped
charge ($\cal Q$) as a function of the length scale $L_{P}$ (or the
temperature $T$). In terms of the Landauer$-$Buttiker conductance,
$G_0 = (2e^2/h)$ $|T_{ee}^0|^2$, using Eq.~\ref{ten}, we obtain
the pumped charge as
\bea {\cal Q} &=& {\cal Q}_{\rm int}-\left({d
\over L_{P}}\right)^{-(2\alpha_{2} -\alpha_{1})} \ \int_0^\tau dt~
I(t)
\nonumber \\
{\rm where}~\nonumber \\
I(t) &=& \non\\
&&
\!\!\!\!\!\!\!\!\!\!\!\!\!\!\!\!\!\!\!\!\!\!\!\!\!\!\!\!\!\!\!\!\!\!\!\!\!
\frac{e}{2 \pi}\frac{G_0 \left[\left(1+2\alpha_{1} \ln
\frac{L_{P}}{d}\right)^\frac{3}{2}\right]\dot{\delta}}{1 + G_0\left[-1 +
\left(1+2\alpha_{1} \ln
\frac{L_{P}}{d}\right)^\frac{3}{2}\left(\frac{d}{L_{P}}\right)^{-(2\alpha_{
2}-\alpha_{1})}\right]} \label{qq} \eea
Here $\delta =\theta + \phi$ and as earlier, $G_0$ is expressed in
unit of $(2e^2/h)$. ${\cal Q}_{\rm int}$ is the integer
contribution of the first term in Eq.~\ref{pc2}.

%
\begin{figure}[htb]
\begin{center}
\includegraphics[width=7cm,height=5.5cm]{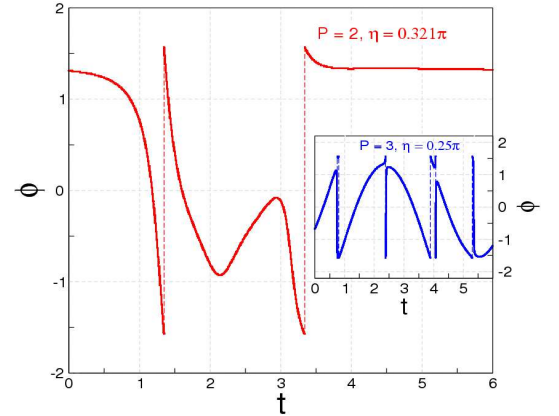}
\caption{The plot shows the variation of the \ar phase $\phi$ with time
$t$, along the pumping contour $C_1$ in the plane of $\Delta_1-\Delta_2$
and the inset shows the variation of the same along the pumping contour
$C_2$ in the plane of $\phi_1-\phi_2$.}
\label{arphase}
\end{center}
\end{figure}

\vskip -1cm
\section{\label{sec:five}{Results and Discussions}}
\vskip -0.5 cm
\begin{enumerate}
\item {\textsf{Pumping in the $\Delta_1-\Delta_2$ plane :}} Here
the pumped charge is obtained by periodically varying the top gate
voltage which controls the Fermi energy of the electrons in the
superconducting region. Hence it amounts to varying $E/\Delta$ for
the two barriers periodically. Just like the double barrier problem,
in this case too we observe resonant transmission of electrons at
discrete values of $E/\Delta$ for fixed values of $\phi_1$ and
$\phi_2$. These discrete values correspond to the existence of
quasi-bound states formed inside the \sdb which are quite different
from their normal double barrier counterpart as they are produced
due to superposition of both electron and hole states and not just
any one of them. In Fig.~\ref{contamp} (left panel), we see sharp
resonance in transmission probability ($|T_{ee}|^{2}$) in the
$\Delta_1-\Delta_2$ plane for $L=1$. We employ the solutions to the
\rg equations (Eq.~\ref{rg}) to obtain the renormalized surface of
transmission in the plane of $\Delta_1-\Delta_2$ for a value of
$L=10$, this is shown in Fig.~\ref{contamp} (right panel). Note that
the \rg flow is such that the transmission increases in the entire
$\Delta_1-\Delta_2$ plane, hence reducing the sharpness of resonance
and resulting in an increase of transmission (conductance) on the
pumping contour $C_1$ giving rise to reduction in  the pumped charge
from its quantized value (see Fig.~\ref{figpumpchargeamp}). From
Fig.~\ref{arphase}, we notice that the \ar phase $\phi$ shows a
total drop in its value by a factor of $2\pi$ during its time
evolution along the contour $C_1$. This drop corresponds to the
quantization of the topological part in the expression for pumped
charge $\cal Q$ (Eq.~\ref{pc2}) to the value of $\ed$.

\begin{figure}[htb]
\begin{center}
\includegraphics[width=7cm,height=5.5cm]{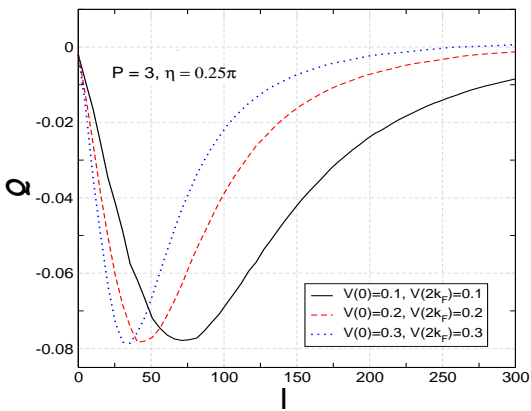}
\caption{Pumped charge $\cal Q$, for pumping in $\phi_1-\phi_2$
plane is shown in the figure as a function of the dimensionless parameter
$l$ where $l=ln(L_{P}/d)$ and $L_{P}$ is either $L_{T}=\hbar v_{F}/k_{B} T$
at zero bias or $L_{V}=\hbar v_{F}/eV$ at zero temperature and $d$ is the
short distance cut-off for the \rg flow. The three curves correspond to
three different values of $V(0)$ and $V(2k_{F})$.}
\label{figpumpchargeph}
\end{center}
\end{figure}
%
\vskip -1.0cm
\item {\textsf{Pumping in the $\phi_1-\phi_2$ plane :}} In
contrast to the previous case, here we obtain two sharp lines of
resonances for the transmission function in the $\phi_1-\phi_2$
plane. Again we observe in Fig.~\ref{contph} that the \rg flow
(Eq.~\ref{rg}) results in reduction of the sharpness of the
resonance.
We consider a pumping contour $C_2$ which encloses parts of both the
resonance lines in the $\phi_1-\phi_2$ plane. The intersection of
the pumping contour $C_2$ with the lines of resonance results in
vanishing of the topological part. This can be seen by observing the
time-evolution of the \ar phase along contour $C_2$ as shown in the
inset of Fig.~\ref{arphase}. In this case the drops are exactly
compensated by corresponding rises in phase $\phi$ by same amount,
leading to a net zero topological contribution to the pumped charge.
Hence for small values of $L_{P}$ (see Fig.~\ref{figpumpchargeph}),
the pumped charge is almost zero. This is because the topological
part is identically zero while the dissipative part is non-zero but
vanishingly small (due to the resonance being very sharp) as the
conductance on most part of the contour is negligible. As we go to
the larger $L_{P}$ values, the pumped charge shows an interesting
non-monotonic behavior, purely coming due to the variation of the
dissipative part.
\end{enumerate}

In conclusion, we show that pumping in the $\Delta_1-\Delta_2$ plane
is much more efficient as opposed to that in $\phi_1-\phi_2$ plane.
We also demonstrate that the quantization of the pumped charge is
lost in $\Delta_1-\Delta_2$ plane if we include correlations due to
proximity effects in the \od \qwd. Although if the barriers are
reflecting then according to \rgd, the system will flow to the
disconnected fixed point ($r=1$) at low temperature. In that case
the sharp transmission resonance would appear in the parameter plane
of back-scattering strength of the first and the second barriers. If
the pumping contour encloses the transmission resonance, then in the
zero temperature limit, the dissipative part of the pumped charge
will become vanishingly small resulting in quantized pumped charge.
So for the \sdb system with small normal reflection, pumped charge
will eventually converge to a quantized value in the zero
temperature limit.

\section*{Acknowledgements}
{We thank Sumathi Rao for many stimulating and useful discussions
and encouragement. We acknowledge use of the Bewoulf cluster at HRI.
The work of SD was supported by the Feinberg Fellowship Programme at
Weizmann, Israel.}




\bibliographystyle{apsrev}

\bibliography{superqpumpref}


\end{document}

%% file: superqpumpmacro.tex
\newcommand\beq{\begin{equation}}
\newcommand\eeq{\end{equation}}

\newcommand\bea{\begin{eqnarray}}
\newcommand\eea{\end{eqnarray}}

\newcommand\bi{\begin{itemize}}
\newcommand\ei{\end{itemize}}

\newcommand\non{\nonumber}

\newcommand\up{\uparrow}
\newcommand\dn{\downarrow}

\newcommand\psiou{\Psi_{O\,\up}^{}}
\newcommand\psiiu{\Psi_{I\,\up}^{}}

\newcommand\psioudg{\Psi_{O\,\up}^\dagger}
\newcommand\psiiudg{\Psi_{I\,\up}^\dagger}

\newcommand\psiod{\Psi_{O\,\dn}^{}}
\newcommand\psiid{\Psi_{I\,\dn}^{}}

\newcommand\psioddg{\Psi_{O\,\dn}^\dagger}
\newcommand\psiiddg{\Psi_{I\,\dn}^\dagger}

\newcommand\hmi{{\mathcal H}_{\textsf{int}}}

\newcommand\odd{1{\textendash}D}

\newcommand\od{1{\textendash}D~}

\newcommand\ie{{\it{i.e.}}}

\newcommand\scbd{{\textsf{SB}}}

\newcommand\scb{{\textsf{SB~}}}

\newcommand\sdbd{{\textsf{SDB}}}

\newcommand\sdb{{\textsf{SDB~}}}

\newcommand\rgd{{\textsf{RG}}}
\newcommand\qwd{{\textsf{QW}}}

\newcommand\nsnsn{{\textsf{NSNSN~}}}

\newcommand\nsnd{{\textsf{NSN}}}

\newcommand\hfd{{\textsf{HF}}}

\newcommand\ard{{\textsf{AR}}}

\newcommand\ed{{\textsf{e}}}

\newcommand\wirg{{\textsf{WIRG~}}}

\newcommand\rg{{\textsf{RG~}}}
\newcommand\qw{{\textsf{QW~}}}

\newcommand\nsn{{\textsf{NSN~}}}

\newcommand\hf{{\textsf{HF~}}}

\newcommand\ar{{\textsf{AR~}}}

\newcommand\bdgd{{\textsf{BdG}}}
\newcommand\bdg{{\textsf{BdG~}}}

\newcommand\dc{{\textsf{DC~}}}

\def\scxone#1{\langle ~#1~ \rangle}

\def\And{{\rm and\ }}

\def\dfrac#1#2{{\displaystyle\frac{#1}{#2}}}

\newif\ifboo \boofalse
